\documentclass{JHEP3}

\usepackage{epsfig}

\newcommand{\be}{\begin{equation}}
\newcommand{\ee}{\end{equation}}

\title{Domain wall filters.\footnote{Talk delivered by H. Neuberger at the
workshop ``Domain Wall Fermions at Ten Years'', March 15-17, 2007, BNL.}}
\author{O. B{\"a}r, \\Institute of Physics, Humboldt University Berlin,\\
Newtonstrasse 15, 12489 Berlin, Germany.\\E-mail: \email{obaer@physik.hu-berlin.de}}
\author{R. Narayanan
\\Department of Physics, Florida International University, Miami,
FL 33199, USA\\E-mail: \email{rajamani.narayanan@fiu.edu}}
\author{ H. Neuberger
\\ Rutgers University, Department of Physics and Astronomy,
Piscataway, NJ 08855, USA\\E-mail: \email
{neuberg@physics.rutgers.edu} }
\author{O. Witzel,  \\Institute of Physics, Humboldt University Berlin,\\
Newtonstrasse 15, 12489 Berlin, Germany.\\E-mail: \email{witzel@physik.hu-berlin.de}}

\abstract {We propose using the extra dimension separating the domain walls
carrying lattice quarks of opposite handedness to gradually filter
out the ultraviolet fluctuations of the gauge fields that are felt
by the fermionic excitations living in the bulk. This generalization of the homogeneous 
domain wall construction has some theoretical features that seem nontrivial.}

\keywords{Chirality, Lattice Gauge Field Theories}

\preprint{HU-EP-07/03 \\ SFB/CPP-07-05 }

\begin{document}

\section{Introduction.}

To explain our basic idea it is best to start using continuum notation
and language.

In Callan and Harvey's original domain wall construction one had fermions
propagating in a full five dimensional gauge background. This situation was
maintained in Kaplan's original proposal, but was early on discarded in
favor of a five dimensional gauge field that is restricted to be independent
of the fifth dimension and whose component in the fifth dimension is zero.

Here we focus only on the vector-like case, with two domain walls, separated
by a fifth dimension of the topology of a circle. The domain walls are
at diametrically opposed points on the circle. On the lattice, one can take
the fermion mass to infinity on one of the semicircles, and effectively
reduce the circle to a segment, with a domain wall at each end.
The Weyl partners that make up a Dirac fermion live separately on the
walls, with small leakage into the ``bulk'', the interior of the fifth
dimension. For a finite separation there is an exponentially small
{\it direct} interaction between the partners, which can be thought of as
an effective vertex of the structure of an exponentially small mass term.
Sometimes, in the lattice field theory literature,
such mass terms are referred to as a ``residual mass''.
Only at infinite separation does one get exactly massless Dirac particles,
with the associated exact chiral symmetry. The latter holds since one
can independently rotate the partners due to the infinite separation between
the walls.

Here we shall re-introduce a dependence of the four components on the
fifth dimension. For the time being, we maintain the fifth component at zero.
However, the five dimensional gauge field is uniquely defined in terms
of a four dimensional background; the dependence on the fifth dimension
is chosen by hand, in a way designed to improve the speed of
convergence to an infinite separation limit.

For our construction we need to introduce the concept of a ``UV'' filter.
This is a term making its appearance in the lattice field theory literature,
but its meaning varies slightly, depending on context. To be specific, we
first define the term in the continuum. A ``UV'' filter ${\cal F}$ will be an operation
that gets as input some four dimensional Euclidean gauge field $A$ and
produces a new four dimensional Euclidean gauge field ${\cal F}(A)$.
We require that ${\cal F}$ and the operation of a four dimensional
gauge transformations of $A$ commute. ${\cal F}(A)$ is smoother
than $A$ in that its the gauge invariant content fluctuates less 
than that of $A$. As an example, we define a filter using
APE smearing:
\be
{\cal F}_{\tau_0}(A)=A(\tau_0),~~~~\partial_\tau A_\mu (\tau) =\partial_{A_\mu} S(A)|_{A=A(\tau)},~~~
A(0)=A
\ee
Here $S$ is some local four dimensional gauge invariant action, like the ordinary
Euclidean Yang Mills action possibly including also higher derivative terms.
$\tau_0$ is a quantity of dimensions length (for separations along the four physical
dimension) squared. $\sqrt{\tau_0}$ is a relatively short distance on the scale of four
dimensional physics. Linearizing the above equation is required to produce, for $\tau >0$ a suppression
for higher momentum modes in $A(\tau)$ in Feynman gauge.

We want to use a UV filter on the gauge fields seen by most of the fermions (bulk)
to hasten the convergence to the infinite wall separation limit. However,
we do not want the UV filter to affect too much the physical fermions, which
live on the walls, because, for example, if we add a mass to make the physical quarks
heavy, we still want the charmonium spectrum to come out right, and this requires
the Coulomb potential to be well represented even at short distances, making UV filtering
undesiarble.

This leads us to introduce a profile in the five dimensional gauge field.
We label the segment connecting the domain walls by the variable $s$, where
$0\le s \le S$ and introduce the profile function $\tau(s)$ which
obeys $\tau(s)=\tau(S-s)$ and $\tau(0)=0$. $\tau(s)$ increased from $0$
gradually to some value $\tau_0$, stays there for a stretch of $s$ given by $l$,
until it reaches the midpoint $s=S/2$, after which it reverses its behavior
in accordance with the above mentioned constraint under reflection. This
constraint is needed to produce an extra symmetry, a parity operation connecting the fermions
bound to the walls. When $l=\infty$ the walls are no longer coupled, and
we have exact chiral symmetry.

As is well known, it is useful to view  $s$ as an Euclidean auxiliary time,
with an associated $s$-dependent Hamiltonian $H(\tau(s))$. Obviously, the $s$-independent
case is easier to analyze. What we gain here is that the gap, $g$, separating
negative and positive energies in $H(s)$ increases in the ``flat'' region
of length $l$. The approach to infinite $l$ is governed essentially by
$e^{-|g|l}$, so it becomes obvious why we want $g$ to be large.
The impact of the
$UV$ fluctuations contained in $A(\tau )$ is such that $|g|$ increases with $\tau$.

There can be other choices of the filter and profile. Also, the
mass term can have a profile, like in the original constructions.
What was described above is close to what we actually implemented
on the computer, in a lattice version of the above.

\section{Definitions.}

We shall label the four dimensional slices by $s=1,2,...,S$; the
physical fermions live on slices $1$ and $S$ and the associated
five dimensional wave functions are localized in the fifth coordinate $s$,
to the vicinities of these walls.

 Following~\cite{almost}, we write the kernel for the five dimensional fermionic
 action in the following form:
\be
\label{fived}
 D=\pmatrix{ C_1^\dagger & B_1 & 0 & 0 & 0 & 0
&\ldots &\ldots & 0 & 0\cr
            B_1        & -C_1 & -1& 0 & 0 & 0 &\ldots &\ldots & 0 & 0\cr
        0   &-1& C_2^\dagger& B_2 & 0 & 0 &\ldots &\ldots & 0 & 0\cr
            0 & 0  &     B_2    &-C_2 &-1 & 0 &\ldots &\ldots & 0 & 0\cr
        0 & 0  & 0 &-1& C_3^\dagger & B_3 &\ldots &\ldots & 0 & 0\cr
            0 & 0  & 0 & 0&  B_3        &-C_3 &\vdots &\ldots & 0 & 0\cr
            \vdots&\vdots&\vdots&\vdots&\vdots&\ddots&\ddots&\ddots
&\vdots&\vdots\cr
            \vdots&\vdots&\vdots&\vdots&\vdots&\vdots&\ddots&\ddots
&\ddots&\vdots\cr
        \ldots&\ldots&\ldots&\ldots&\ldots&\ldots&\ldots&\ddots
&\ddots&\ldots\cr
        0 & 0 &  0 & 0& 0         & 0 &\ldots &\ldots & B_S &-C_S\cr}
\ee

The matrix $D$ is of size $2k\times 2k$ where the entries are
$q\times q$ blocks and $k=S$ and $q=2NL^4$, where the gauge
group is taken as $SU(N)$. The factor $2N$ counts spinorial and
gauge group indices.

The matrices $B_s$ and $C_s$ are dependent on the gauge background
defined by the collection of link matrices $U^s_\mu (x)$. These
matrices are of dimension $N\times N$. $\mu$ labels the positive
$4$ directions on a hypercubic lattice and  $U^s_\mu (x)$ is the
unitary matrix associated with a link that points from the site
$x$ in the $\hat \mu$-direction on slice s
\begin{eqnarray}
& ( C_s )_{x \alpha i, y \beta j}  ={1\over 2} \sum_{\mu=1}^{d}
\sigma_\mu^{\alpha\beta} [\delta_{y,x+\hat\mu} (U^s_\mu (x) )_{ij}
- \delta_{x,y+\hat\mu} (U_\mu^{s \dagger} (y))_{ij}] \equiv
\sum_{\mu=1}^{d} \sigma_\mu^{\alpha\beta} ( W^s_\mu )_{xi,yj}
\cr & ( {B_s}_0 )_{x \alpha i, y \beta j}  = {1\over 2}
\delta_{\alpha\beta} \sum_{\mu =1}^{d} [2\delta_{xy} \delta_{ij} -
\delta_{y,x+\hat\mu} (U^s_\mu (x) )_{ij} - \delta_{x,y+\hat\mu}
(U_\mu^{s \dagger} (y))_{ij} ]\cr &( B_s )_{x \alpha i, y \beta j}
 =( {B_s}_0 )_{x \alpha i, y \beta j} + M^s_0 \delta_{x\alpha
i,y\beta j}\cr
\end{eqnarray}

The indices $\alpha,\beta$ label spinor indices in the range $1$
to $2$. The indices $i,j$ label color in the range $1$ to $2$. The
Euclidean $4 \times 4$ Dirac matrices $\gamma_\mu$ are taken in
the Weyl basis where their form is
\be
\gamma_\mu = \pmatrix {0&\sigma_\mu \cr \sigma_\mu^\dagger & 0\cr}
\ee

As long as
$M^s_0 > 0$ the matrices $B_s$ are positive definite due to the
unitarity of the link variables. To make almost massless quarks on
the lattice one also wants $|M^s_0|<1$. 

It is convenient to introduce the $2q \times 2q$ matrix $\Gamma_5$
representing the regular $\gamma_5$ matrix on spinorial indices
and unit action on all other indices. In terms of $q\times q$
blocks we have
\be
\Gamma_5 =\pmatrix{1&0\cr0&-1} \ee

\section{The effective four dimensional fermion action.}

Following the method of ~\cite{almost} we get

\be
\det D = (-)^{qk} (\prod_{s=1}^S\det B_s ) \det \left [
{{1-\Gamma_5}\over 2} -{\bf T}_l {{1+\Gamma_5}\over 2}\right ] \ee

Define

\be
T_s \equiv e^{-H_s} = \pmatrix {{1\over B_s} & {1\over B_s} C_s
\cr C_s^\dagger {1\over B_s} & C_s^\dagger {1\over B_s} C_s +B_s
\cr} \ee

For any $s$, $\det T_s =1$.

${\bf T}_l$ is a symmetric product of $T_s^{-1}$
factors. By definition, gauge fields and mass parameters $M_0$ labeled
by $s$ and $S-s$ are identical.

\be
{\bf T}_l = R^\dagger T_r^{-(l+2)} R \ee

The $R,R^\dagger$ factors are complex matrices (neither unitary,
nor hermitian) representing ``ramps'' and are given by:

\be
R=T^{-1}_{r} T^{-1}_{r-1} .. T^{-1}_1,~~~~  R^\dagger =T^{-1}_S
T^{-1}_{S-1} .. T^{-1}_{S-r} \ee

In between the ``ramps'' we have a uniform ``plateau'' with identical 
transfer matrix factors. The total number of slices is $k=l+2r$

\be
\det D = (-)^{qk} (\prod_{s=1}^S \det B_s ) \det \left [1 +{\bf
T_l}^{-1} \right ] \det \left [ {{1+\Gamma_5 \frac{1-{\bf
T}_l}{1+{\bf T}_l} }\over 2} \right ] \ee

The same derivation for Pauli Villars (PV) fields should give
a PV determinant made out of the first two factors in the formula above.
Dividing the two expressions, leaves us with the last factor
as representing the almost massless Dirac fermion.

Thus, the operator
\be
E_l=\frac{1-{\bf
T}_l}{1+{\bf T}_l}
\ee
carries the $l$ dependence and its spectrum governs the approach to
the chiral limit $l=\infty$.

Unlike in the homogeneous case, we are so far unable to 
prove that $E_l$
indeed has a limit, $E_\infty$, as $l$ tends to 
infinity. If we assume that such
a limit exits, the rate of convergence would be controlled by the
eigenvalue of ${\bf T}_l$ that is closest to unity as $l$ becomes
very large.

Physically, using effective Lagrangian intuition, one would expect $E_\infty$ 
to exist for most four dimensional gauge fields. Simulations with the five dimensional
action could practically resolve this question.

\section{How to test the method.}

Before proceeding to a five dimensional implementation one would like 
to find by numerical methods the lowest or few lowest eigenvalues of the matrix
${\bf T}_l+{\bf T}_l^{-1}$ but this is tough because the condition number of ${\bf T}_l$ 
rapidly becomes too large as the number of slices increases.
A possible trick to get around this goes as follows:

It is easy to check that the spectrum of
\be
B_S=\pmatrix {0& A_1 & 0 &\dots &\dots & 0\cr
              0& 0 & A_2 &\dots &\dots & 0\cr
          \vdots&\vdots&\vdots&\vdots&\vdots &\vdots \cr
          0&0&\dots&\dots&\dots&A_{S-1}\cr
          A_S&\vdots&\vdots&\vdots&\vdots&0\cr}
\ee
is given by $\lambda_{n,k}=\rho_n^{1/S} e^{\frac{\imath \phi_n}{S}}
e^{\frac{2\pi\imath k}{S}}$, with $k=0,1,2,...,S-1$ where $\lambda_n=\rho_n e^{\imath\phi_n}$
are the eigenvalues of 
$\Pi_S\equiv A_1....A_s$. 
In our application, $\Pi_S$ is hermitian and positive, so $\phi_n=0$.
From the relation $|\lambda_{n,k}|=\rho_n^{1/S}$ we see that we would get directly
the quantity we expect to have a finite limit as $l\to\infty$. Writing
$\rho_n\equiv e^{S\kappa_n}$, with a weak $S$-dependence in $\kappa_n$ suppressed,
we can find the dominating $\kappa_n$ using a routine based on a 
restarted Arnoldi procedure looking for the eigenvalues of $B_S-1$ of smallest
magnitude ~\cite{matrix}. 

In some sense, it might be surprising that this relatively simple generalization
of the homogeneous domain wall setup is substantially more difficult to
analyze theoretically, with or without numerical tools. Since $E_\infty$ would
provide a new variant of an overlap operator, the issue might be of interest 
separately from the question of increased numerical efficiency of QCD domain
wall simulations. 

\section{Further directions.}

One could adapt to the lattice a definition of
$A(s)$ based on simple interpolation.
This would allow for an adiabatic limit, with an almost homogeneous extra dimension  
where theoretical issues would be under better control, setting aside the question
how far one can deviate from this limit.

\acknowledgments

RN acknowledges partial support by the NSF under grant number
PHY-055375 and also partial support from Jefferson
Lab. The Thomas Jefferson National Accelerator Facility (Jefferson
Lab) is operated by the Southeastern Universities Research
Association (SURA) under DOE contract DE-AC05-84ER40150. HN 
acknowledges partial support by the DOE under grant number
DE-FG02-01ER41165 at Rutgers University and also 
thanks the Alexander von Humboldt foundation for an award that made this
work possible. HN also would like to thank the Physics Department at Humboldt
University for the warm hospitality extended.
OW acknowledges support by the DFG within the SFB/TR 9 
``Computational Particle Physics''.


\begin{thebibliography}{99}
\bibitem{almost} H. Neuberger, Phys. Rev. D57, 5417 (1998)
\bibitem{matrix}  ``Matrix Computations'', Gene H. Golub, Charles F. Van Loan, Johns
Hopkins (1993). 
\end{thebibliography}
\end{document}